\documentclass[twocolumn]{aastex631}

\shorttitle{Jupiter as a Bipolytrope}
\shortauthors{Kadam K.}
\graphicspath{{./}{figures/}}

\begin{document}

\title{Jupiter as a Rotating Bipolytrope}

\author[0000-0001-8718-6407]{Kundan Kadam}
\email{kkadam@uwo.ca}
\affiliation{Department of Physics and Astronomy, University of Western Ontario, London, Ontario, N6A 3K7, Canada}



\begin{abstract}
Polytropes have long been used to model a wide variety of astrophysical objects.
A bipolytrope (composite polytrope) may be used for bodies with a distinct core-envelope structure.
In this short paper, I demonstrate that a rotating bipolytrope is a reasonable approximation for Jovian interior.
Similar models may be used to probe rotating exoplanets to gain an intuitive understanding of their internal structure.  

\end{abstract}



\section{Introduction} \label{sec:intro}


Polytropic equation of state (EoS), wherein the pressure is decoupled from the temperature, is widely used to gain an understanding of the interiors of astrophysical objects \citep{Chandrasekhar33}.
Polytropes have been used to investigate the structure and evolution of stars and planets, moons in the solar system, compact objects, globular clusters as well as molecular clouds \citep[e.g.][]{Horedt04}.  
For bodies having a core with significantly distinct physical properties, a bipolytrope may be used \citep{Henrich1941,Murphy1983}.
A bipolytrope can have two different polytropic indices for the core and envelope which allows for different EoS's.
The difference in compositions or a phase transition can also be taken into account.
It is well known that rotation can significantly affect the structure and evolution of most astrophysical objects \citep[e.g.][]{Meynet1997}.
In \cite{Kadam+16}, we extended Hachisu's self-consistent field (SCF) method for obtaining rapidly rotating, gravitationally bound structures in equilibrium to bipolytropes \citep{Hachisu1986a}.
In this exercise, I consider a particular bipolytropic model of Jupiter \citep{Criss2015} and demonstrate that it can be improved upon by adding rotation. 
Such models can be used to probe interiors of exoplanets if their rotational properties are observed.

\section{Jupiter as a Bipolytrope}

A polytropic EoS assumes a power-law relationship between the pressure of the gas and its density such that
\begin{equation}
P = \kappa \rho^{1+\frac{1}{n}},    
\end{equation}
where $n$ is the polytropic index, $\kappa$ is the polytropic constant.
The ratio of molecular weights, $\alpha=\mu_{\rm c}/\mu_{\rm e}$, equals the density jump across the core-envelope interface. 
The subscripts ``${\rm c}$" and ``${\rm e}$" denote the values for the core and the envelope, respectively. 
The fractional radius is defined as the ratio of the equatorial radius of the core to that of the whole configuration (i.e., $q = \mathcal{R}_{\rm c}/\mathcal{R}$).
The bipolytropic self-consistent field (BSCF) method obtains the equilibrium configuration of a rotating bipolytrope by solving Poisson's equation for gravitational potential and hydrostatic equation iteratively, until the convergence of the virial error \citep{Kadam+16}. 

A polytropic index of unity can be assumed to represent Jovian interior \citep{Stevenson1982}.
However, Jupiter contains a gaseous, helium depleted mantle and a condensed metallic hydrogen core, in addition to a much smaller rocky core \citep{Hubbard2013}. 
\cite{Criss2015} approximated the Jovian structure as a bipolytrope with the metallic hydrogen core as an incompressible solid ($n_{\rm c}=0$), and a gaseous, diatomic mantle ($n_{\rm e}=2.5$). 
The model assumed a core-envelope interface at $q = 0.8$ and a large discontinuity at this location corresponding to $\alpha \approx 36.7$. 
Henceforth, this model is called CH. 
The total normalized moment of inertia (or radius of gyration, ${k_{\rm t}}^2$) for a configuration is defined as
\begin{equation}
     I_{\rm norm}=I / \mathcal{M} \mathcal{R}^2,
\end{equation}
in the conventional notation.
\cite{Criss2015} show that $I_{\rm norm}$ for an $n=1$ polytrope is inconsistent with the observed value of 0.254 for Jupiter \citep{NASA2014}, and model CH is a better approximation. 

\begin{figure}
\centering
\includegraphics[width=3in]{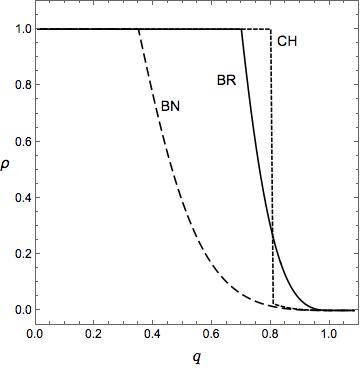}
\caption{The normalized equatorial density profile of Jupiter as a bipolytrope for models CH \citep{Criss2015}, BN (non-rotating bipolytrope) and BR (rotating bipolytrope). 
}
\label{fig:jupiter}
\end{figure}

Model CH has two major disadvantages. First, there is no justification of an abrupt and large jump ($\alpha \approx 36.7$) at the interface between core and the envelope.
\cite{Nellis2000} showed that the phase transition between molecular to monoatomic metallic hydrogen is continuous so Jupiter should have no distinct core-envelope boundary.
The second drawback of model CH is the infinitesimal contribution of the envelope to the structural parameters. 
The gaseous envelope contributes only about $1.7\%$ to the total $I_{\rm norm}$ and $0.6 \%$ to the total mass, essentially making the envelope fit irrelevant, as long as its contribution remains small enough (see Fig. \ref{fig:jupiter}). 

Both of these issues can be addressed by considering a rotating bipolytropic model.
Consider a model with $n_{\rm c}=0.0001$ (a low value of $n_{\rm c}$, instead of zero due to the implementation), $n_{\rm e}=2.5$ and with the discontinuity at the interface eliminated by setting $\alpha=1$.
With no rotation (model BN), the normalized moment of inertia equals 0.25 when $q=0.34$. 
This size of the condensed core is much smaller than that expected for Jupiter, as metallization of fluid hydrogen is supposed to occur much closer to the surface, possibly as high as 0.9 $\mathcal{R}$ \citep{Nellis2000}. 

Jupiter is a fast rotating planet with flattening, $f$, defined as 
$f = {\rm (a-b)/a}$
where ${\rm a}$ is the equatorial radius and ${\rm b}$ is the polar radius. 
We add rotation to model BN (model BR) by setting an axis ratio of 0.93, corresponding to the observed flattening value of 0.064 \citep{NASA2014}. 
With rotation, the core fractional radius goes back up to $q=0.69$ in order to make the normalized moment of inertia 0.25. 
This implies that one cannot account for a large core and observed $I_{\rm norm}$ simultaneously without rotation.
The normalized density profiles for all three models (CH, BN and BR) are shown in Fig. \ref{fig:jupiter}. 

The BR model makes a few predictions about the physical properties of the gas giant. 
For the average density of Jupiter to be $1.33$ $\rm gcm^{-3}$ as derived from the planet's mass, we need to set the maximum density to $2.72$ $\rm gcm^{-3}$. 
With this maximum density and the angular velocity obtained with model BR, we calculate the rotational period of 8.21 hours. 
This period is a lower limit as the actual central density will be greater than that of the condensed metallic hydrogen core, since the planet is assumed to have a much denser Earth-sized core composed of heavy metals \citep{Hubbard2013}.

The external multipole moments or the zonal harmonics coefficients $J_l$ of an axisymmetric gravitational potential are given by
\begin{equation}
J_l=-\frac{2 \pi}{\mathcal{M} \mathcal{R}^l}   \int_0^{\pi} \int_0^{r_{\rho=0}}  \rho(r,\theta) P_l(\cos \theta)  r^{l+2} \sin{\theta} \mathrm{d}r \mathrm{d}\theta,
\end{equation} 
where $P_l(\cos\theta)$ are the Legendre functions.
The zonal coefficients are measured by the effects of the planet's gravity on a spacecraft \citep[e.g. Juno,][]{Bolton+17}, and they mainly reflect the rotational distortion of the density distribution of the planet.
The gravitational coefficients for model BR are 
$J_2=1.29 \times 10^{-2}(12.1)$,
$J_4=-3.74 \times 10^{-4}(36.3)$ and 
$J_6=1.36 \times 10^{-5}(73.9)$, where the quantity in the bracket shows the percentage error as compared to the measurements. 
Although model BR is relatively crude approximation with polytropic EoS and only two-layer structure, the zonal harmonic coefficients match reasonably well to the observed values.
More elaborate models have already been published for the Jovian interior.
For example, the equilibrium solutions of \cite{Hubbard2013}, which reproduce the zonal coefficients to a much better accuracy, use as many as 512 uniformly rotating concentric Maclaurin spheroids. 
However, the model obtained with the BSCF method attempts to strike a balance between complexity and intuitive understanding.

\section{Discussion}
In this letter, I demonstrated that a rotating bipolytrope is a better model for Jovian interior as compared to analogous non-rotating models. 
In the future, the BSCF method may be modified to incorporate the rocky (or icy) core, either as a point mass or another polytrope, as well as the differential rotation of the planet.
Similar models can be used to probe the internal structures of exoplanets such as the fast-spinning, young gas giant $\beta$ Pictoris b \citep{Snellen2014}.

\bibliography{references_file} 

\begin{thebibliography}{}
\expandafter\ifx\csname natexlab\endcsname\relax\def\natexlab#1{#1}\fi

\bibitem[{{Bolton} {et~al.}(2017){Bolton}, {Adriani}, {Adumitroaie}, {Allison},
  {Anderson}, {Atreya}, {Bloxham}, {Brown}, {Connerney}, {DeJong}, {Folkner},
  {Gautier}, {Grassi}, {Gulkis}, {Guillot}, {Hansen}, {Hubbard}, {Iess},
  {Ingersoll}, {Janssen}, {Jorgensen}, {Kaspi}, {Levin}, {Li}, {Lunine},
  {Miguel}, {Mura}, {Orton}, {Owen}, {Ravine}, {Smith}, {Steffes}, {Stone},
  {Stevenson}, {Thorne}, {Waite}, {Durante}, {Ebert}, {Greathouse}, {Hue},
  {Parisi}, {Szalay}, \& {Wilson}}]{Bolton+17}
{Bolton}, S.~J., {Adriani}, A., {Adumitroaie}, V., {et~al.} 2017, Science, 356,
  821

\bibitem[{{Chandrasekhar}(1933)}]{Chandrasekhar33}
{Chandrasekhar}, S. 1933, \mnras, 93, 390

\bibitem[{{Criss} \& {Hofmeister}(2015)}]{Criss2015}
{Criss}, R.~E., \& {Hofmeister}, A.~M. 2015, New Astronomy, 36, 26

\bibitem[{{Hachisu}(1986)}]{Hachisu1986a}
{Hachisu}, I. 1986, \apjs, 61, 479

\bibitem[{{Henrich} \& {Chandrasekhar}(1941)}]{Henrich1941}
{Henrich}, L.~R., \& {Chandrasekhar}, S. 1941, \apj, 94, 525

\bibitem[{{Horedt}(2004)}]{Horedt04}
{Horedt}, G.~P. 2004, {Polytropes - Applications in Astrophysics and Related
  Fields}, Vol. 306, doi:10.1007/978-1-4020-2351-4

\bibitem[{{Hubbard}(2013)}]{Hubbard2013}
{Hubbard}, W.~B. 2013, \apj, 768, 43

\bibitem[{{Kadam} {et~al.}(2016){Kadam}, {Motl}, {Frank}, {Clayton}, \&
  {Marcello}}]{Kadam+16}
{Kadam}, K., {Motl}, P.~M., {Frank}, J., {Clayton}, G.~C., \& {Marcello}, D.~C.
  2016, \mnras, 462, 2237

\bibitem[{{Meynet} \& {Maeder}(1997)}]{Meynet1997}
{Meynet}, G., \& {Maeder}, A. 1997, \aap, 321, 465

\bibitem[{{Murphy}(1983)}]{Murphy1983}
{Murphy}, J.~O. 1983, \pasa, 5, 175

\bibitem[{NASA(2014)}]{NASA2014}
NASA. 2014, http://nssdc.gsfc.nasa.gov/planetary/factsheet/-jupiterfact.html

\bibitem[{{Nellis}(2000)}]{Nellis2000}
{Nellis}, W.~J. 2000, \planss, 48, 671

\bibitem[{{Snellen} {et~al.}(2014){Snellen}, {Brandl}, {de Kok}, {Brogi},
  {Birkby}, \& {Schwarz}}]{Snellen2014}
{Snellen}, I.~A.~G., {Brandl}, B.~R., {de Kok}, R.~J., {et~al.} 2014, \nat,
  509, 63

\bibitem[{{Stevenson}(1982)}]{Stevenson1982}
{Stevenson}, D.~J. 1982, Annual Review of Earth and Planetary Sciences, 10, 257

\end{thebibliography}
\bibliographystyle{aasjournal}

\end{document}